\documentclass[12pt,a4paper,oneside]{article}
\usepackage{ifpdf}
\usepackage{a4wide}
\usepackage{amsmath}
\usepackage{graphicx}
\usepackage{rotating}
\usepackage[numbers,sort&compress]{natbib} 
\usepackage{amssymb}
\usepackage[latin9]{inputenc}
\begin{document}

\textheight 21.0cm
\textwidth 16cm
\sloppy
\oddsidemargin 0.0cm \evensidemargin 0.0cm
\topmargin 0.0cm

%\begin{document}
\setlength{\parskip}{0.45cm}
\setlength{\baselineskip}{0.75cm}

%XXXXXXXXXXXXXXXXXXXXXXXXXXXXXXXXXXXXXX

%XXXXXXXXXXXXXXXXXXXXXXXXXXXXXXXXXXXXXX

\begin{titlepage}
\setlength{\parskip}{0.25cm}
\setlength{\baselineskip}{0.25cm}
\begin{flushright}
DO-TH 10/06\\
ZU-TH 05/10\\
\vspace{0.2cm}
March 2010
\end{flushright}
\vspace{1.0cm}
\begin{center}
\Large
{\bf On the charged current neutrino-nucleon total\\ 
cross section at high energies}
\vspace{1.5cm}

\large
M.\ Gl\"uck$^1$, P.\ Jimenez-Delgado$^2$,  and E.\ Reya$^1$
\vspace{1.0cm}

\normalsize
$^1${\it Universit\"{a}t Dortmund, Institut f\"{u}r Physik}\\
{\it D-44221 Dortmund, Germany} \\
\vspace{0.5cm}

$^2${\it Universit\"{a}t Z\"{u}rich, Institut f\"ur Theoretische
Physik}\\
{\it CH-8057, Z\"{u}rich, Switzerland}\\

\vspace{1.5cm}
\end{center}

\begin{abstract}
\noindent 
We evaluate the charged current neutrino-nucleon total cross section
up to neutrino energies of $10^{12}$ GeV in the LO and NLO of perturbative
QCD within the framework of two different factorization schemes.
The numerical implications of some inconsistent QCD calculations are
illustrated.
\end{abstract}
\end{titlepage}

%XXXXXXXXXXXXXXXXXXXXXXXXXXXXXXXXXXXXXXXXXXXXXXXXXXXXXXXXXXXXXXXXX
%MAIN PART
%XXXXXXXXXXXXXXXXXXXXXXXXXXXXXXXXXXXXXXXXXXXXXXXXXXXXXXXXXXXXXXXX

In a recent publication \cite{ref1}, utilizing our latest 
next-to-leading order (NLO) parton distribution functions
(pdfs) in the  `fixed flavor number scheme' (FFNS) \cite{ref2}
and  `variable flavor number scheme' (VFNS) \cite{ref3}, some
conclusions concerning the charged current (CC) neutrino-nucleon
total cross sections at high energies were drawn which deserve
a closer examination.  For this purpose we shall follow the standard
formalism as outlined in \cite{ref4} which operates within the 
framework of the ${\overline{\rm MS}}$ factorization scheme compatible
with the ${\overline{\rm MS}}$ distributions in \cite{ref2,ref3}.

The perturbative stability is reexamined in Fig.\ 1(a) which 
presents $K_{\rm FFNS}$ and $K_{\rm VFNS}$ where 
$K \equiv \sigma_{\rm NLO}^{\nu N}/\sigma_{\rm LO}^{\nu N}$. 
These results differ from the ones presented in Figs.~5 and 6
of \cite{ref1} where the NLO pdfs were inconsistently also folded
with the leading order (LO) sub-cross sections.  
The effects of this inconsistency
are demonstrated by comparing the inconsistent results in Fig.~1(b) 
with the results in Fig.~1(a). These
inconsistent results in Fig.~1(b) are similar to the ones in
Fig.~6 of \cite{ref1}. It should be furthermore noted that, 
contrary to the procedure in \cite{ref1}, our NLO FFNS predictions
are {\em not} treated via the ACOT formula (10) in \cite{ref1} 
but rather according to the aforementioned ${\overline{\rm MS}}$
prescriptions \cite{ref4} which is the consistent way to regularize
the NLO mass singularities within the  ${\overline{\rm MS}}$ 
factorization scheme utilized in \cite{ref2,ref3}.  Clearly, once
pdfs have been extracted from experiment by employing one specific
factorization scheme, they must {\em{not}} be used for calculations
based on different schemes.  Anyway, the difference between the 
results in Fig.~1(a) and the inconsistent ones in Fig.~1(b) is
mainly caused by the inconsistent use of NLO pdfs for LO calculations
and far less caused by the mismatch of schemes \cite{ref1}. 
For completeness we also show in Fig.~1 the predictions based on
our previous FFNS GRV98 pdfs \cite{ref5}, which differ very little
from the ones based on our more recent FFNS GJR pdfs \cite{ref2}.

A comparison of the VFNS and FFNS predictions in LO and NLO is
presented in Fig.~2 showing the corresponding 
$\sigma_{\rm VFNS}^{\nu N} / \sigma_{\rm FFNS}^{\nu N}$ ratios.
(Note that the VFNS pdfs in \cite{ref3} have been generated from
the ones  in the FFNS \cite{ref2} where the full heavy quark mass
dependence has been taken into account.) 
The VFNS expectations are similar for the CTEQ pdfs \cite{ref6}
with a K-factor somewhat closer to 1 at highest energies, i.e.,
about 0.7 at $E_{\nu}=10^{12}$ GeV instead of 0.6 for our VFNS
pdfs \cite{ref3} according to Fig.~1(a). The difference between
the VFNS and FFNS amounts to at most about 20\% at highest energies
which is a typical theoretical uncertainty due to different
(possible) choices of factorization schemes.  It should be
emphasized that the VFNS is by no means superior to the FFNS at 
large scales ($Q^2\gg m_{c,b}^2$) as commonly claimed.  
This is due to the fact that
in the VFNS the heavy quark flavors ($c$, $b$, and eventually $t$)
also become massless partons within the nucleon with distributions
obtained from massless evolutions starting at the
``thresholds'' $Q^2=m_{c,b,t}^2$.  Thus it is an additional
{\em assumption}, rather than a theoretical necessity, that these
massless ``heavy'' quark pdfs are relevant
asymptotically and that they correctly describe the asymptotic
behavior of deep inelastic structure functions and cross sections
for large scales.
In fact if perturbative stability is considered as a selective
criterion, then Fig.~1(a) speaks in favor of the FFNS at very 
high neutrino energies, i.e. at
$E_{\nu}$ \raisebox{-0.1cm} {$\stackrel{>}{\sim}$} $10^8$ GeV,
where, according to Fig.~2, the NLO VFNS and FFNS predictions
differ considerably.

The predicted total cross sections as calculated at NLO in the 
FFNS can be inferred from Fig.~3 which can be combined with Fig.1(a)
and/or Fig.~2 whenever desired.  We also depict in Fig.~3 the 
individual contributions from the CC subprocesses of the light
(\mbox{$W^+ d\to u$}, $W^+ s\to u$, $W^+\bar{u}\to \bar{d}$, etc.)  
and heavy
quarks, with the charm contribution deriving from the $W^+s\to c$
and $W^+d\to c$ transitions and the ${\cal{O}}(\alpha_s)$ 
contributions \makebox{$W^+g\to c\bar{s}$}, $W^+s'\to gc$, etc., where
$s'_{\nu N}\equiv|V_{cs}|^2s +|V_{cd}|^2(d+u)/2$, as described,
for \mbox{example}, in \cite{ref4,ref7} for the  ${\overline{\rm MS}}$ 
factorization scheme, with $V_{ij}$ denoting the standard CKM matrix 
elements. The fact that the heavy $t\bar{b}$ FFNS
contribution has so far only been calculated in LO 
($W^+g\to t\bar{b}$) is of minor importance, since the $c\bar{s}$
sector dominates over the much heavier $t\bar{b}$ one.  For
consistency reasons we calculated this fully massive 
($m_{b,t}\neq 0$) contribution always using the LO gluon distribution.
This is immaterial at 
$E$ \raisebox{-0.1cm} {$\stackrel{<}{\sim}$} $10^8$ GeV 
where the $t\bar{b}$ contribution to $\sigma_{\rm tot}^{\nu N}$ is
negligible while at higher energies this amounts to an uncertainty
of only a few percent due to the fact that the NLO massive $t\bar{b}$
contribution would amount to an ${\cal{O}}(\alpha_s)$ correction to 
the LO
contribution of about 20\%.

To summarize, we have demonstrated that fully consistent LO and 
NLO QCD calculations of CC neutrino--nucleon total cross sections
within the framework of two different factorization schemes 
(FFNS, VFNS) yield rather stable perturbative results even at 
ultra-high neutrino energies of about $10^{12}$ GeV.  The FFNS
results turn out to be more stable than the VFNS ones at highest
energies.  The difference between the FFNS and VFNS results amounts
to at most about 20\% at highest energies of $10^{10}$ to $10^{12}$
GeV, which is a typical theoretical uncertainty due to different
(possible) choices of factorization schemes.

\newpage
%%%%%%%%%%%%%%%%%%%%%%%%%%%%%%%%%%%%%%%%%%%%%%%%%%%%%%%%%%%%%%%%%%%%%%%%%%%%%
%%%%%%%%%%%%%%%%%%%%%%%%%%%%%%%%Fig. 1%%%%%%%%%%%%%%%%%%%%%%%%%%%%%%%%%%%%%%%%%%%
\begin{figure}
\begin{center}
\ifpdf
\includegraphics[width=14.0cm]{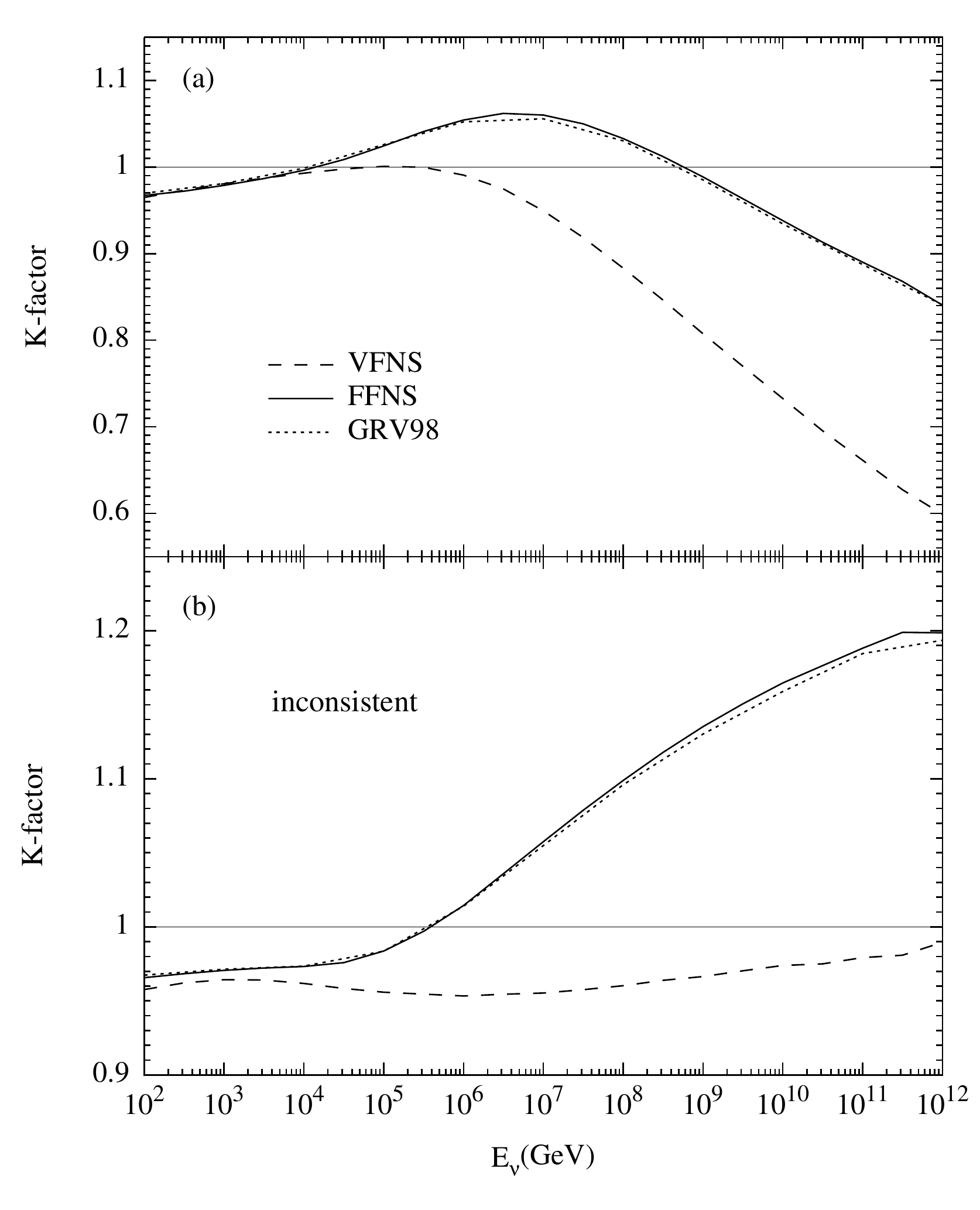}
\fi
\caption{(a) Predictions for the $K\equiv \sigma_{\rm NLO}^{\nu N}/
\sigma_{\rm LO}^{\nu N}$
factors of the CC cross sections in two different factorization schemes,
using the FFNS GJR \cite{ref2} and GRV98 \cite{ref5}  pdfs, and the VFNS pdfs
of \cite{ref3}.  The inconsistent K-factors in (b) have been evaluated
using always the same NLO pdfs also at LO as was done in \cite{ref1}.}
\end{center}
\end{figure}
\clearpage
%%%%%%%%%%%%%%%%%%%%%%%%%%%%%%%%Fig. 2%%%%%%%%%%%%%%%%%%%%%%%%%%%%%%%%%%%%%%%%%
\begin{figure}
\begin{center}
\ifpdf
\includegraphics[width=14.0cm]{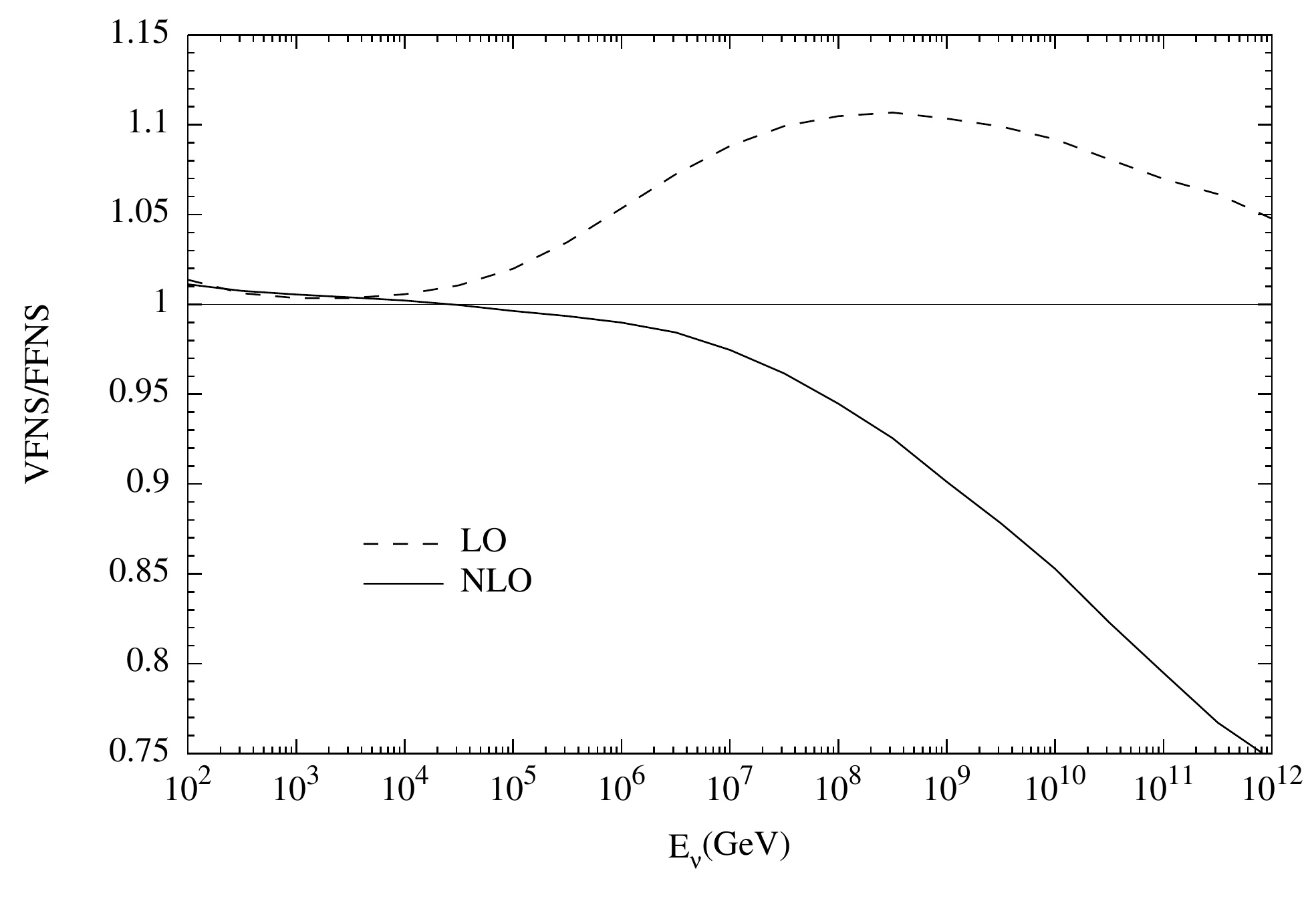}
\fi
\caption{The ratios of CC $\nu N$ cross sections as calculated in the
VFNS \cite{ref3} and FFNS \cite{ref2} at LO and NLO.  The ratios are
similar if the VFNS CTEQ6 pdfs \cite{ref6} are used instead.} 
\end{center}
\end{figure}
\clearpage
%%%%%%%%%%%%%%%%%%%%%%%%%%%%%%%%%Fig. 3%%%%%%%%%%%%%%%%%%%%%%%%%%%%%%%%%%%%%%%%%%
\begin{figure}
\begin{center}
\ifpdf
\includegraphics[width=14.0cm]{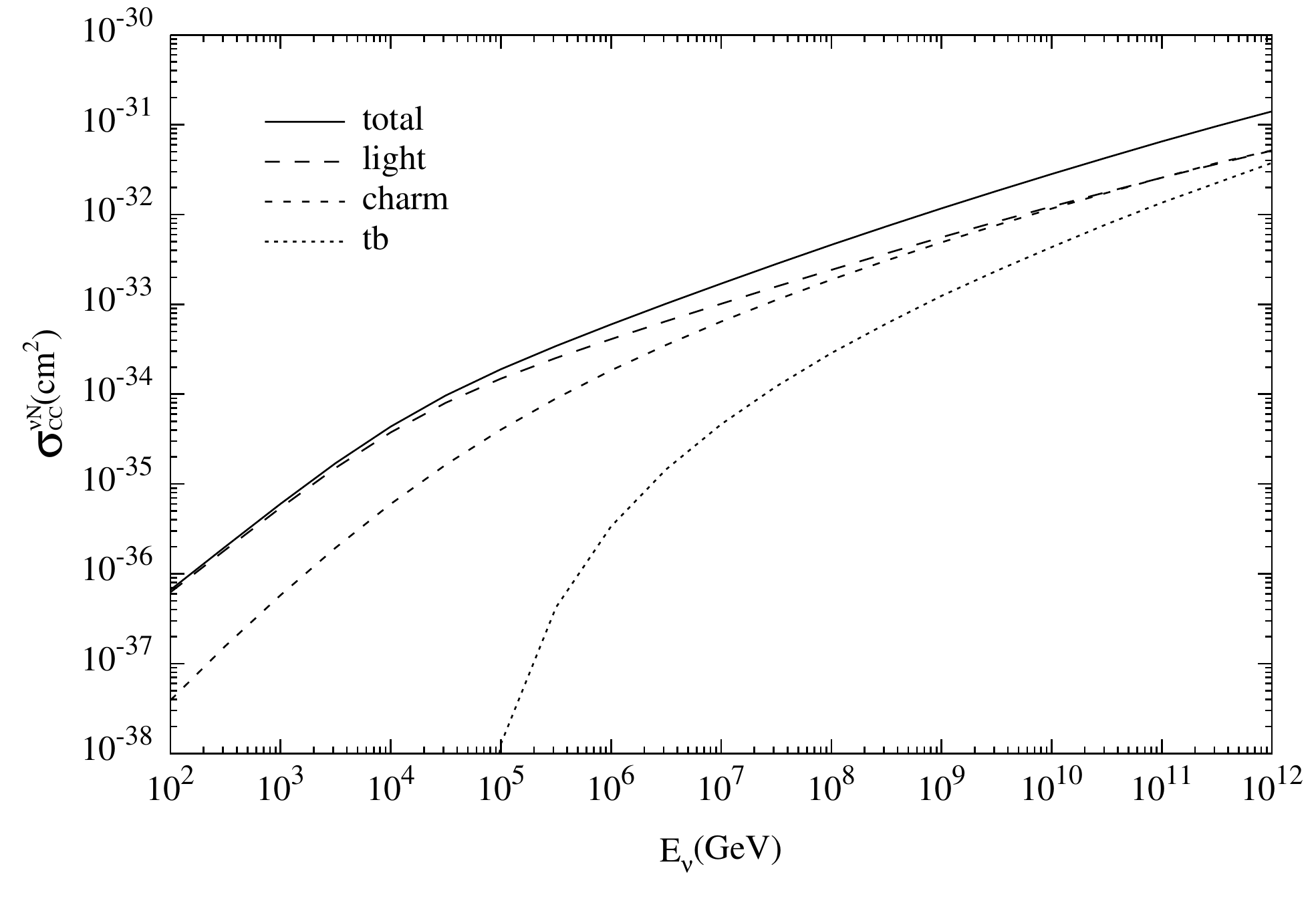}
\fi
\caption{The total CC $\nu N$ cross section as calculated at NLO in
the FFNS using the GJR pdfs \cite{ref2}. The individual contributions
due to light ($d\to u$, $s\to u$, $\bar{u}\to \bar{d}$, etc.) and
charm ($s\to c$, $d\to c$, etc.) quark CC transitions are shown as
well.  The fully massive $tb$ contribution has so far been calculated
only in LO according to the subprocess $W^+g\to t\bar{b}$ which, for
consistency reasons, has to be folded also with the LO gluon 
distribution.}
\end{center}
\end{figure}
\end{document}